\documentclass[a4paper,11pt]{article}
\pdfoutput=1 

\usepackage{jheppub} 

\usepackage[T1]{fontenc} 

\newcommand{\arctanh}{\text{arctanh}}

\newcommand{\mA}{\mathcal{A}}

\newcommand{\mC}{\mathcal{C}}
\newcommand{\mD}{\mathcal{D}}

\newcommand{\mS}{\mathcal{S}}
\newcommand{\mL}{\mathcal{L}}

\newcommand{\mN}{\mathcal{N}}
\newcommand{\mO}{\mathcal{O}}

\newcommand{\mQ}{\mathcal{Q}}
\newcommand{\mV}{\mathcal{V}}

\newcommand{\mfC}{\mathfrak{C}}

\newcommand{\Poincare}{Poincar\'e }

\newcommand{\be}{\begin{eqnarray}}
\newcommand{\ee}{\end{eqnarray}}

\def\>{\rangle}
\def\<{\langle}

\def\tr{\hbox{Tr}}

\newcommand{\tcr}{\textcolor{red}}

\newcommand{\executeiffilenewer}[3]{%
	\ifnum\pdfstrcmp{\pdffilemoddate{#1}}%
	{\pdffilemoddate{#2}}>0%
	{\immediate\write18{#3}}\fi%
}

\graphicspath{{pics/}}

\title{\boldmath A complexity/fidelity susceptibility $g$-theorem for AdS$_3$/BCFT$_2$}

\author{Mario Flory}

\affiliation{Institute of Physics, Jagiellonian University, \\
	\L{}ojasiewicza 11, 30-348 Krak\'ow, Poland}

\emailAdd{mflory@th.if.uj.edu.pl}

\abstract{We use a recently proposed holographic Kondo model as a well-understood example of AdS/boundary CFT (BCFT) duality and show explicitly that in this model the bulk volume decreases along the RG flow. We then obtain a proof that this volume loss is indeed a generic feature of AdS/BCFT models of the type proposed by Takayanagi in 2011. According to recent proposals holographically relating bulk volume to such quantities as complexity or fidelity susceptibility in the dual field theory, this suggests the existence of a complexity or fidelity susceptibility analogue of the Affleck-Ludwig $g$-theorem, which famously states the decrease of boundary entropy along the RG flow of a BCFT. We comment on this possibility.   
}

\begin{document} 
\maketitle
\flushbottom

\section{Introduction}
\label{sec::Intro}

The AdS/CFT correspondence \cite{Maldacena:1997re,Gubser:1998bc,Witten:1998qj} suggests that certain conformal field theories (CFTs) can be holographically dual to gravity theories with asymptotically AdS spacetimes. The AdS/CFT correspondence can hence be used as a tool to gain understanding of various field theory phenomena, usually at strong coupling, by phrasing the problem at hand in terms of the dual gravitational theory. Amongst many other things, topics to which AdS/CFT methods have been applied in the past were RG flows (see e.g.~\cite{Freedman:1999gp,deBoer:1999tgo}), holographic superconductors \cite{Gubser:2008px,Hartnoll:2008vx,Amado:2009ts} and \textit{boundary CFTs (BCFTs)}, see \cite{Chiodaroli:2009yw,Takayanagi:2011zk,Fujita:2011fp,Nozaki:2012qd,Estes:2015jha,Miao:2017gyt,Chu:2017aab,Astaneh:2017ghi} for a cursory overview over different types of AdS/BCFT models proposed in the literature. 

In descriptions of the AdS/CFT correspondence, one will often encounter the term \textit{holographic dictionary}, describing the idea of a, real or imagined, list of quantities defined either on the bulk (AdS) or field theory (CFT) side that are mapped to each other via the correspondence. An example of a well-known entry into this dictionary is the holographic \textit{entanglement entropy} formula \cite{Ryu:2006bv,Ryu:2006ef}
\begin{align}
S=\frac{\mA}{4 G_N},
\end{align}
where the entanglement entropy $S$ of a certain subregion is a CFT quantity, while $G_N$ is the (bulk) Newton constant and $\mA$ is the area of an extremal co-dimension two surface in the bulk (AdS) spacetime. Recently, there were two independent proposals for what might be the field theory dual to the volumes of certain extremal co-dimension \textit{one} hypersurfaces.

Firstly, ideas relating to \textit{computational complexity} seem to have entered holography in discussions concerning the \textit{firewall paradox} \cite{Almheiri:2012rt} in the works \cite{Harlow:2013tf,Susskind:2013aaa,Susskind:2014rva}, where a connection between complexity and bulk geometry was envisioned. Following \cite{Susskind:2014rva} (see also \cite{Stanford:2014jda,Susskind:2014jwa,Susskind:2014yaa,Susskind:2015toa}), we will define \textit{computational complexity} as the minimal number of simple unitary operations that have to be carried out by a quantum computer in order to implement a given unitary operation on a simple initial state, or to create a given state from a simple initial state. Based on the findings of \cite{Susskind:2013aaa,Susskind:2014rva}, it was then suggested in \cite{Susskind:2014moa,Stanford:2014jda,Susskind:2014jwa,Susskind:2014yaa,Susskind:2015toa} that holographically, the complexity $\mC$ of the field theory state should 
be measured by the volumes $\mV$ of certain spacelike extremal co-dimension one bulk hypersurfaces, i.e.\footnote{As discussed e.g.~in \cite{Stanford:2014jda,Susskind:2014jwa,Brown:2015bva,Brown:2015lvg}, for dimensional reasons the length scale $L$ has to be introduced into equation \eqref{16complexity}, which leads to a certain arbitrariness in the choice of the scale in this definition. In the following, we will use equation \eqref{16complexity} with $L$ being the AdS radius as the definition of complexity, assuming an order one factor of proportionality between the left and right hand side.}
\begin{align}
\mC\propto \frac{\mV}{LG_N}.
\label{16complexity}
\end{align}
It was later argued in \cite{Brown:2015bva,Brown:2015lvg} that the computational 
complexity $\mC$ should more accurately be calculated from the integral of the bulk action over a certain (co-dimension zero) bulk region, the Wheeler-DeWitt patch. However, the simple approximation formula \eqref{16complexity} has continued to attract interest in the holography community, see \cite{Alishahiha:2015rta,Barbon:2015ria,Momeni:2016ekm,Mazhari:2016yng,Ben-Ami:2016qex,Momeni:2016ira,Roy:2017kha}. We will hence work with \eqref{16complexity} in this paper, and comment on the action proposal of \cite{Brown:2015bva,Brown:2015lvg} (see also \cite{Chemissany:2016qqq,Cai:2016xho,Brown:2016wib,Lehner:2016vdi,Couch:2016exn,Roberts:2016hpo,Chapman:2016hwi,Huang:2016fks,Carmi:2016wjl,Pan:2016ecg,Reynolds:2016rvl,Brown:2017jil,Kim:2017lrw,Zhao:2017iul} for further work in this direction) again in section \ref{sec::conc}.

Secondly, in \cite{MIyaji:2015mia} it was proposed that the volume $\mV$ of an extremal spacelike co-dimension one hypersurface should be approximately dual to a quantity $G_{\lambda\lambda}$ called \textit{quantum information metric} or \textit{fidelity susceptibility} according to the formula
\begin{align}
G_{\lambda\lambda}=n_d \frac{\mV}{L^{d}},
\label{fidsus}
\end{align}
where $n_d$ is an order one factor, $L$ is the AdS radius and $d$ determines the dimension such that the AdS space is $d+1$ dimensional. For two normalised states $\left|\psi(\lambda)\right>$ and $\left|\psi(\lambda+\delta\lambda)\right>$ belonging to a one-parameter family of states, $G_{\lambda\lambda}$ is defined as\footnote{As the left hand side is bounded from above by one and $\delta\lambda$ can have any sign, there cannot be a term of order $\delta\lambda$.}
\begin{align}
|\left<\psi(\lambda)\middle|\psi(\lambda+\delta\lambda)\right>|=1-G_{\lambda\lambda}\delta\lambda^2+\mO(\delta\lambda^3)
\end{align}
and measures the distance between the two states, hence the name quantum information metric. The name fidelity susceptibility derives from the fact that $|\left<\psi(\lambda)\middle|\psi(\lambda+\delta\lambda)\right>|$ is called the fidelity. As discussed in \cite{MIyaji:2015mia} (see also \cite{Bak:2015jxd,Trivella:2016brw}), $G_{\lambda\lambda}$ can be holographically calculated when the two states $\left|\psi(\lambda)\right>$ and $\left|\psi(\lambda+\delta\lambda)\right>$ are the ground states of a theory allowing for a holographic dual, and when the difference $\delta\lambda$ is the result of a perturbation of the Hamiltonian by $\delta\lambda \cdot\hat{O}$ with an exactly marginal operator $\hat{O}$. The correct bulk spacetime dual to this field theory problem is a so called \textit{Janus solution} \cite{Bak:2003jk,Bak:2007jm}, but in \cite{MIyaji:2015mia} it was shown that this geometry may be approximated by a simpler spacetime with a probe defect brane embedded into it. This then leads to the formula \eqref{fidsus}. See \cite{Alishahiha:2015rta,Momeni:2016qfv,Mazhari:2016yng,Sinamuli:2016rms,Banerjee:2017qti} for further results using prescription \eqref{fidsus}.

The structure of this paper is as follows: In section \ref{sec::Kondo} we will first recapitulate the framework proposed in \cite{Takayanagi:2011zk,Fujita:2011fp,Nozaki:2012qd} to build AdS/BCFT models, and then we will present a specific example of a model of this type: The holographic Kondo model of 
\cite{Erdmenger:2013dpa,Erdmenger:2014xya,OBannon:2015cqy,Erdmenger:2015spo,Erdmenger:2015xpq,Erdmenger:2016vud,Erdmenger:2016jjg,Erdmenger:2016msd}. We will present this model, show some of the most important results derived from it so far, and explain why it is a particularly well-behaved model of the type \cite{Takayanagi:2011zk,Fujita:2011fp,Nozaki:2012qd}. Section \ref{sec::volumeloss} will then be denoted to showing that in this Kondo model, the bulk volume $\mV$ of the $t=0$ slice of the spacetime decreases monotonically as the temperature is lowered. In section \ref{sec::theorem} we will prove that under certain physical assumptions, this behaviour is indeed generic in AdS$_3$/BCFT$_2$ models. As we will discuss in section \ref{sec::conc}, according to the proposals \eqref{16complexity} and \eqref{fidsus}, this implies the existence of a complexity and/or fidelity susceptibility analogue of the Affleck-Ludwig $g$-theorem \cite{Affleck:1991tk} for holographic BCFTs. This theorem famously implies that, for BCFTs, the \textit{boundary entropy} $\ln g(T)$ is a monotonic function of the temperature $T$,
\begin{align}
T \cdot\frac{\partial }{\partial T}\ln(g)\geq 0,
\label{lngT}
\end{align}
where lowering the temperature can be interpreted as going from the UV to the IR \cite{Affleck:1991tk,PhysRevB.48.7297,Yamaguchi:2002pa,Friedan:2003yc}\footnote{\label{bulkflow}It should be pointed out that the $g$-theorem only holds when the BCFT undergoes an RG flow of its boundary, but remains critical otherwise, i.e.~while the boundary entropy $\ln(g)$ changes, the central charge $c$ of the BCFT is assumed not to change. Otherwise, the change of $\ln(g)$ may have any sign, see \cite{Green:2007wr} and the discussion at the end of \cite{Bak:2016rpn}. }.


\section{Review: AdS/BCFT and a Holographic Kondo Model}
\label{sec::Kondo}

\subsection{AdS/BCFT}
\label{sec::AdSBCFT}

\begin{figure}[htb]
	\centering
	\def\svgwidth{0.35\columnwidth}
	\executeiffilenewer{NMQP3.svg}{NMQP3.pdf}%
	{inkscape -z -D --file=NMQP3.svg %
		--export-pdf=NMQP3.pdf --export-latex}%
	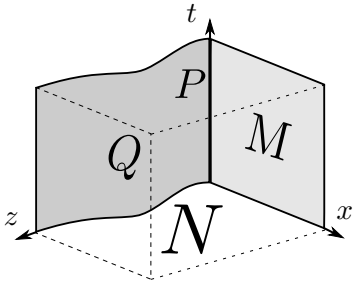%

	\caption{Setup for the holographic description of a BCFT according to \cite{Takayanagi:2011zk}. The asymptotically 
		AdS bulk spacetime $N$ has the conformal boundary $M$ and additional boundary $Q$. 
		The defect $P$ is the intersection of $M$ and $Q$. We have used standard coordinates $t,x,z$ as in \eqref{BTZ}, where $t,x$ are boundary directions and $z$ increases into the bulk. The figure is taken from \cite{Erdmenger:2014xya}.
	}
	\label{fig::NMQP}
\end{figure}

A BCFT is a CFT that lives on a space that has a boundary, such as the half-plane for example. Often \textit{defect-} and \textit{interface-CFTs} can also be equivalently formulated as BCFTs, so we will not distinguish these terms in the following. There are several ways to study BCFTs holographically\cite{Chiodaroli:2009yw,Takayanagi:2011zk,Fujita:2011fp,Nozaki:2012qd,Estes:2015jha,Miao:2017gyt,Chu:2017aab,Astaneh:2017ghi}, and in the rest of this paper we will work in a bottom-up framework proposed by Takayanagi and others in \cite{Takayanagi:2011zk,Fujita:2011fp,Nozaki:2012qd}. The underlying idea of this proposal is very simple: In standard AdS/CFT, we work with a \textit{bulk spacetime} $N$. This spacetime, being asymptotically AdS, has a \textit{conformal boundary} $M$, on which conventionally the holographically dual field theory is interpreted to live. In order to describe a BCFT, this space $M$ then has to have a boundary $P$ itself, which in the following we will refer to as the \textit{defect} in order to avoid confusion. Holographically, in the framework of \cite{Takayanagi:2011zk,Fujita:2011fp,Nozaki:2012qd} this defect should then be extended into the bulk spacetime $M$ by a co-dimension one hypersurface $Q$, which will will refer to as \textit{brane}. Hence both $M$ and $Q$ will in a sense be boundaries of the bulk spacetime $N$, but with the important difference that $M$ will be the asymptotic boundary, on which boundary and counter-terms have to be imposed, while $Q$ will be considered to be a part of the classical bulk description of the dual theory. Especially, we will allow for arbitrary classical matter fields to live in the worldvolume of $Q$, holographically describing the degrees of freedom of the BCFT restricted to $P$. See figure \ref{fig::NMQP}. Furthermore, the boundary condition for the bulk metric $g_{\mu\nu}$ at $Q$ will be chosen to be a Neumann boundary condition in the terminology of \cite{Takayanagi:2011zk}\footnote{See also \cite{Krishnan:2016mcj} for a deeper discussion of what can be called a Neumann boundary condition in general relativity.}, i.e.~the induced metric $\gamma_{ij}$ on $Q$ will be allowed to fluctuate. The bulk-action for an AdS$_{d+1}$/BCFT$_d$ model of this type will then read \cite{Takayanagi:2011zk,Fujita:2011fp,Nozaki:2012qd}\footnote{The sign in front of the extrinsic curvature term depends on the chosen convention convention. As in \cite{Erdmenger:2014xya, Erdmenger:2015spo}, we will choose the normal vector of $Q$ to be pointing \textit{inwards}. This yields the signs as in equation \eqref{Action},\eqref{EOM}.}
\begin{align}
\mS =\frac{1}{2\kappa_N^2}\int_N d^{d+1}x\, 
\sqrt{-g}\left(R-2\Lambda+2\kappa_N^2\mL_N\right)
-\frac{1}{\kappa_N^2}\int_Q d^d x\, \sqrt{-\gamma} 
\left( K-\kappa_N^2\mL_Q\right)
+ S_{c.t.}^{(M,P)}.
\label{Action}
\end{align}
Here, $\kappa_N^2$ is related to Newton's $G_N$ constant by $\kappa_N^2=8\pi G_N$, $\mL_N$ is the Lagrangian of matter fields in the bulk $N$, $K$ is the extrinsic curvature tensor on $Q$ and $\mL_Q$ is the Lagrangian for matter fields living on the worldvolume of the hypersurface $Q$. $S_{c.t.}^{(M,P)}$ are boundary and counter terms defined on $M$ and $P$. When calculating the equations of motion from this action, apart from the usual equations for the fields living in $N$, we obtain the following equation determining the embedding of $Q$ into $N$:
\begin{align}
K_{ij} - \gamma_{ij} K = -\kappa_N^2\,S_{ij}.
\label{EOM}
\end{align}
Here $K_{ij}$ is the extrinsic curvature tensor of $Q$, $\gamma_{ij}$ is the induced metric on $Q$ and $S_{ij}$ is the energy-momentum tensor derived from the fields $\mL_Q$. As also explained in \cite{Erdmenger:2014xya}, the similarity between \eqref{EOM} and the \textit{Israel junction conditions} \cite{Israel:1966rt}, describing under which conditions two spacetimes can be glued together along a common boundary surface, is no coincidence. Simply speaking, when assuming the gluing carried out via the Israel junction conditions to be mirror symmetric with respect to the gluing surface, the Israel junction conditions exactly reproduce \eqref{EOM} up to a factor of $\frac{1}{2}$ in front of the stress energy tensor. In fact, the Israel junction conditions can also be derived from an action principle ansatz of the form \eqref{Action}, but with the bulk $N$ split into two components $N_\pm$ sharing a common boundary surface $Q$ \cite{Hayward:1990tz,Hayward:1993my}. The difference between such a two sided approach and the one-sided approach \eqref{EOM} would be interpreted  as the difference between a holographic model of a defect CFT (DCFT) and a genuine BCFT. For the remainder of this paper, this distinction will not be relevant, and our results will be applicable to both holographic DCFTs and BCFTS as long as they are described by the equations of motion \eqref{EOM}.

The concrete geometry of the model will be determined as follows. For a given  bulk metric $g_{\mu\nu}$ on $N$, and for a fixed $P$, the embedding of $Q$ into $N$ can be parametrised in terms of embedding functions, for which \eqref{EOM} serves as equation of motion. This equation \eqref{EOM}, together with the Einstein equations for $g_{\mu\nu}$ and the equations of motion for all matter fields living on $N$ and $Q$ then form a coupled system of differential equations that has to be solved. As a simpler example, consider the case where, as in the rest of this paper, we set $d=2$ and specifically look at situations where both the ambient spacetime $N$ and the embedding of $Q$ into $N$ are static. This means that $g_{\mu\nu}$ is supposed to be static and $P$, by staticity, is just a straight line on the boundary at fixed boundary coordinate. Furthermore, let us for simplicity assume $\mL_N=0$. The bulk spacetime $g_{\mu\nu}$ is then a static vacuum solution of $2+1$ dimensional Einstein gravity, which we recognize as the BTZ black hole \cite{Banados:1992gq,Banados:1992wn} 
\begin{align}
ds^2=g_{\mu\nu} d x^\mu  d x^\nu
=\frac{1}{z^2}\left(-h(z)  d t^2 + \frac{ d z^2}{h(z)}+ d 
x^2\right),
\label{BTZ}
\end{align}
with $h(z)=1-z^2/z_H^2$. For simplicity, we have set the AdS radius $L=1$. Without loss of generality, we can fix the position of $P$ to be $x\equiv0$, and the embedding of $Q$ into $N$ is then described by an \textit{embedding profile} $x_+(z)$ (no $t$ dependence due to staticity). We can then calculate the extrinsic curvature tensor $K_{ij}$ in terms of $x_{+}$,
\begin{align}
K_{ij}&=\frac{1}{z ^2 z_H^2 \left(z_H^2-z ^2\right) \sqrt{1+\left(1-\frac{z ^2}{z_H^2}\right) x_+'(z )^2}}
\label{Ktensor}
\\
&\times
\left(
\begin{array}{cc}
-\left(z_H^2-z ^2\right)^2 x_+'(z ) & 0 \\
0 & z  z_H^2 \left(z ^2-z_H^2\right) x_+''(z )+z_H^4 x_+'(z )+\left(z_H^2-z ^2\right)^2 x_+'(z )^3 \\
\end{array}
\right),
\end{align}
and treat $x_+(z)$ as a dynamical field of our model with the equation of motion \eqref{EOM}. For quantities like the induced metric $\gamma_{ij}$ or the extrinsic curvature tensor \eqref{Ktensor}, the indices $i,j$ run over the coordinates $t,z$. One benefit of this approach is that for static AdS$_3$/BCFT$_2$, $x_+(z)$ only depends on one coordinate and hence \eqref{EOM} is a set of ODEs. This allows for a number of elegant exact solutions to be obtained \cite{Erdmenger:2014xya}.

\subsection{A holographic Kondo model}
\label{sec::Kondomodel}

In this and the next section, we will revisit a specific AdS/BCFT model inspired by holographic studies of the \textit{Kondo effect.} See \cite{Kondo01071964} for the original source on the Kondo effect, \cite{2001cond.mat..4100K,Hewson:2009} for a modern 
perspective and \cite{doi:10.1143/JPSJ.74.1} for a brief historical overview. 

This effect has first been observed by measuring the resistivity of metal probes with a low concentration of impurity atoms as a function of temperature. For example, when investigating the resistivity of a gold probe with dilute iron impurities it is found that as the temperature is lowered, the resistivity first attains a minimum at a certain temperature and then \textit{increases} \cite{MacDonald161}. It was quickly 
realised that the phenomenon had to be due to the interaction of conduction electrons with the localised single magnetic impurities \cite{Kondo01071964,doi:10.1143/JPSJ.74.1}. A perturbative second order calculation by Jun Kondo in \cite{Kondo01071964} then explained the rise of the resistivity at low temperatures as a consequence of the spin-spin interaction between impurities and electrons, but also predicted an unphysical \textit{divergence} of the resistivity in the zero temperature limit. This signifies a breakdown of the perturbation theory below a certain temperature $T_K$, the \textit{Kondo temperature} \cite{Kondo01071964,2001cond.mat..4100K,Hewson:2009,RevModPhys.47.773}. The desire to understand the 
correct behaviour of these impurity systems at temperatures below the Kondo temperature $T_K$, the \textit{Kondo 	problem} \cite{Hewson:2009}, inspired the application and development of a variety of different physical methods \cite{2001cond.mat..4100K}, including renormalisation group methods \cite{0022-3719-3-12-008,RevModPhys.47.773}. The modern understanding of the solution of this problem is that at low temperatures the impurity is screened from the rest of the system by conduction electrons that form the \textit{Kondo screening cloud}.

Holographic models of the Kondo effect or qualitatively Kondo like physics were presented in \cite{Harrison:2011fs,Benincasa:2011zu,Benincasa:2012wu,Erdmenger:2013dpa}, and we will specifically focus on a bottom-up model proposed in \cite{Erdmenger:2013dpa} and further studied in
\cite{Erdmenger:2014xya,OBannon:2015cqy,Erdmenger:2015spo,Erdmenger:2015xpq,Erdmenger:2016vud,Erdmenger:2016jjg,Erdmenger:2016msd}, referring the reader to these works for all but the most relevant details. The action of this model is of the form \eqref{Action}\footnote{Strictly speaking, the action is in the configuration appropriate for \textit{defect} CFTs, i.e.~with the bulk spacetime $N$ being divided in two parts $N_\pm$ to the left and to the right of $Q$. As we will be mostly working on the level of the equations of motion assuming symmetry with respect to the defect, this will not make a difference as explained above. See \cite{Erdmenger:2014xya,Erdmenger:2015spo} for more details on this matter.} with $d=2$\footnote{Due to a $s$-wave reduction, the Kondo effect can be described by a $1+1$-dimensional BCFT \cite{Affleck:1990zd}. Hence the bulk in this model is $2+1$ dimensional. } and\footnote{As this model was originally studied as a \textit{two}-sided \textit{defect} CFT in \cite{Erdmenger:2013dpa,Erdmenger:2015spo}, we introduce an additional factor $1/2$ in \eqref{Kondobraneaction} compared to \cite{Erdmenger:2013dpa,Erdmenger:2014xya,Erdmenger:2015spo}. This ensures that when solving the equations of motion, the right hand side of \eqref{EOM} will have the same magnitude as the analogous equation solved in \cite{Erdmenger:2014xya,Erdmenger:2015spo}. }
\begin{align}
\int_N d^{3}x\, 
\sqrt{-g}\mL_N&=-\frac{\mN}{4\pi}\int \tr\left(A\wedge 
d A+\frac{2}{3}A\wedge A\wedge A\right),
\label{CSaction}
\\
\int_Q d^2 x\, \sqrt{-\gamma}\mL_Q&=- \frac{\mN}{2} \int d^2 x\, 
\sqrt{-\gamma}\,\left(\frac{1}{4}f_{mn}f^{mn} + 
\gamma^{mn}(\mD_m \Phi)^{\dagger} (\mD_n\Phi) +V(\Phi\Phi^{\dagger})\right)\, ,
\label{Kondobraneaction}
\\
\mD_m\Phi &\equiv \partial_m\Phi +i q A_m\Phi-i q a_m\Phi \, .
\label{16covDer}
\end{align} 
Here $\mN$ is a normalisation factor that was discussed in \cite{Erdmenger:2013dpa,Erdmenger:2015spo}. We see that the only fields living in the entire bulk spacetime are the bulk metric $g_{\mu\nu}$ and a Chern-Simons gauge field $A$. The meaning of this CS field for the interpretation of the model \eqref{CSaction}-\eqref{Kondobraneaction} is elaborated upon in \cite{Erdmenger:2013dpa,OBannon:2015cqy}, however as this field effectively decouples from the other fields \cite{Erdmenger:2013dpa,OBannon:2015cqy,Erdmenger:2015spo}, and as in this work we are more interested in this model as a generic well-behaved toy model of the type of equation \eqref{Action} than as a concrete Kondo model, we will ignore this field from now on. The fields on the hypersurface $Q$ constitute something similar to a holographic superconductor in $AdS_2$, with the gauge group of the $a$ field chosen to be $U(1)$. In the simplest incarnation of this model, the potential of the charged scalar is chosen to be a pure mass term
\begin{align}
V(\Phi\Phi^{\dagger})=M^2\Phi\Phi^{\dagger},
\label{Mass}
\end{align}
with the mass tuned to the Breitenlohner-Freedman bound \cite{Breitenlohner:1982jf} appropriate for a charged scalar in AdS$_2$ \cite{Iqbal:2011aj}. The bulk spacetime is chosen to be a BTZ \textit{black brane} \eqref{BTZ} where the boundary coordinate $x$ is assumed to be decompactified, $x\in(-\infty,+\infty)$. 

Even in the case including backreaction ($\kappa_N\neq0$), it is generically possible to find an analytic solution to the equations of motion of this model in which $\Phi(z)=0$ but $a_m(z)\neq0$ for every $z$, and we refer to this as the \textit{uncondensed} or \textit{normal phase}. Choosing the gauge $a_z(z)=0$, this solution reads \cite{Erdmenger:2015spo}
\begin{align}
a_t 	&= \frac{\mfC}{z_H} \cosh (s) 
\left(\cosh(s)-\sqrt{(z_H/z)^2+\sinh^2 (s)}\right) \, ,
\label{normalgauge}
\\
x_+(z)
&=-z_H
\,\arctanh\left(\frac{\sinh(s)}{\sqrt{(z_H/z)^2+\sinh^2(s)}}\right)
\, .
\label{backgroundSolutionGaugeAndEmbedding}
\end{align}
with $\mfC^2=-\frac{1}{2}f^{mn}f_{mn}$ the constant electric flux of the gauge field and
\begin{align}
\tanh\left(s\right)=\frac{1}{4}\kappa_N^2 \mN \mfC^2 \, .
\label{backgroundSolutionGeodesicLength}
\end{align}
This solution has the special feature that the energy momentum tensor on $Q$ takes the form $S_{ij}=const.\times \gamma_{ij}$, which is known as a \textit{constant tension} model. Such constant tension models are especially important in AdS/BCFT as they can be solved analytically in many cases and describe RG fixed points, see \cite{Azeyanagi:2007qj,Takayanagi:2011zk,Fujita:2011fp,Nozaki:2012qd,Erdmenger:2014xya}. Expanding \eqref{normalgauge} near the boundary, we find
\begin{align}
a_t\sim -\frac{\mfC L^2\cosh(s)}{z}+\frac{\mfC L^2\cosh(s)^2}{z_H}+...\equiv \frac{\mQ}{z}+\mu_c+...
\label{Qmuc}
\end{align}

Of course, there are also non-trivial solutions with $\phi(z)\neq0$. In this case, the asymptotic expansion of $a_t$ will still take the form 
\begin{align}
a_t\sim \frac{\mQ}{z}+\mu+...
\label{at}
\end{align}
where we fix $\mQ$ and let the chemical potential $\mu$ vary. This leads to a very important point: As said above, we use the metric \eqref{BTZ} with $x\in(-\infty,+\infty)$. This means that we can set
\begin{align}
\tilde{z} = z/z_H,\quad 
\tilde{x} = x/z_H,\quad
\tilde{t}=t/z_H,\quad
\tilde{\phi} =\phi,\quad
\tilde{a}_t = a_t \,z_H,\quad
\tilde{x}_+ = x_+/z_H,\quad
\text{etc.}
\label{tilde}
\end{align}
Leaving away the tildes later on, this means that we can effectively set $z_H=1$,\footnote{In the standard BTZ black hole with periodic identification $x\sim x+2\pi$, such a rescaling would violate this periodicity condition and is hence forbidden. 
} which is precisely what we will do in section \ref{sec::volumeloss}. The rescaling \eqref{tilde} has the effect on \eqref{at} that 
\begin{align}
\tilde{a}_t \sim \frac{\tilde{\mQ}}{\tilde{z}} + \tilde{\mu} +...
\end{align}
with $\mQ=\tilde{\mQ}$ and
\begin{align}
\tilde{\mu}=\mu\cdot z_H=\frac{\mu}{2\pi T}.
\label{tildemu}
\end{align}
We hence see that the chemical potential $\mu$ sets a scale to compare the temperature $T$ to, with the only relevant physical combination being \eqref{tildemu}. In the numerics of \cite{Erdmenger:2015spo}, to be presented in section \ref{sec::volumeloss}, fixing $z_H=1$ and increasing $\mu$ is hence physically equivalent to keeping $\mu$ fixed and decreasing the temperature $T$ below a critical temperature $T_c$. $T/T_c$ is then a function of \eqref{tildemu} and can be used to label the different solutions. This is similar to the situation in holographic superconductors \cite{Amado:2009ts}. In the following, we will mostly think about the Kondo model in terms of a temperature being lowered for fixed $\mu$, as this is the more realistic viewpoint when comparing to experimental studies of the Kondo effect. From now on, we will leave the tildes away on all quantities.     

Increasing $\mu$ above the value $\mu_c$ defined in \eqref{Qmuc}, respectively lowering $T/T_c$ below $1$, we find that the scalar field $\Phi(z)=\phi(z)e^{i\psi(z)}$ in the bulk condenses, and obtains an asymptotic expansion (gauge-fixing $\psi(z)=0$) \cite{Erdmenger:2013dpa}
\begin{align}
\phi(z) 
\sim\alpha \sqrt{z}\log(z)+\beta \sqrt{z}+\ldots
\end{align}
where, due to some peculiarities of the holographic Kondo model explained in detail in \cite{Erdmenger:2013dpa,OBannon:2015cqy}, the appropriate boundary conditions are of the \textit{double trace} type \cite{Witten:2001ua,Berkooz:2002ug}
\begin{align}
\alpha = \kappa \beta,
\end{align} 
with the \textit{Kondo coupling} $\kappa$. Qualitatively, the Kondo model is similar to a holographic superconductor \cite{Gubser:2008px,Hartnoll:2008vx,Amado:2009ts}: At $T=T_c$, the normal phase described above becomes unstable, and at $T<T_c$ the charged scalar field attains a non-vanishing profile, the \textit{condensed} or \textit{broken phase}. Holographically, this is interpreted as the formation of the \textit{Kondo cloud} in the field theory side \cite{Erdmenger:2013dpa}.\footnote{\label{FN::largeN} The holographic Kondo model hence describes the Kondo effect in terms of a phase transition. Due to the Coleman-Mermin-Wagner theorem, this is possible for the real world Kondo effect only in the large $N$ limit, where $SU(N)$ is the spin-group of the magnetic impurity \cite{PhysRevLett.17.1133,Coleman1973,coleman2015introduction}. } Lowering $T/T_c$ (or equivalently increasing $\mu$), the Kondo model then experiences an RG flow from the UV fixed point described by the constant tension solution \eqref{normalgauge}-\eqref{backgroundSolutionGaugeAndEmbedding} towards an IR fixed point \cite{Erdmenger:2013dpa}.

Irrespectively of whether the model presented in this section is an accurate \textit{Kondo} model (see \cite{Erdmenger:2013dpa,Erdmenger:2014xya,OBannon:2015cqy,Erdmenger:2015spo,Erdmenger:2015xpq,Erdmenger:2016vud,Erdmenger:2016jjg,Erdmenger:2016msd} for discussions of this question), it is important to point out that, in its own right, this model is a very interesting AdS/BCFT toy model following the proposal of  \cite{Takayanagi:2011zk,Fujita:2011fp,Nozaki:2012qd}. The reasons for that are as follows:

\begin{itemize}
	\item The model is non-trivial, i.e.~it has non-trivial matter content $\mL_Q\neq const.$, in contrast to the simpler constant tension models. 
	\item The model is well-behaved, in the sense that the matter fields $\mL_Q$ satisfy (or violate) different energy-conditions just in such a way as is phenomenologically necessary for a holographic Kondo model \cite{Erdmenger:2014xya,Erdmenger:2015spo}.  
	\item The model is qualitatively well-understood, as the various energy conditions constrain the possible geometries \cite{Erdmenger:2014xya,Erdmenger:2015spo}.  
	\item The model nicely displays a physical boundary RG flow. As we will discuss in section \ref{sec::Simp}, it is for example possible to explicitly calculate the \textit{boundary entropy} and verify that the Affleck-Ludwig $g$-theorem \eqref{lngT} is satisfied \cite{Erdmenger:2015spo}. 
\end{itemize}	

As the proposals \eqref{16complexity} and \eqref{fidsus} suggest that certain bulk volumes will have an interesting physical interpretation in terms of field theory quantities, we will in the next section calculate the loss of volume that occurs in the holographic Kondo model along the RG flow. 

\section{Volume loss in the Kondo model}
\label{sec::volumeloss}

\subsection{Numerical backreaction and calculation of volume loss}
\label{sec::complexity}

In \cite{Erdmenger:2015spo}, we numerically obtained the embedding profiles $x_+(z)$ solving \eqref{EOM} in the Kondo model \eqref{CSaction}-\eqref{16covDer} with the effective parameter-choices $\kappa_N=\mN=q=1$ and $\mfC=1/2$. These choices are slightly different from the original incarnation of the holographic Kondo model \cite{Erdmenger:2013dpa}, but they allow for non-trivial backreaction and stable numerics. Furthermore, in \cite{Erdmenger:2015spo} we argued that the geometry of the backreacted solutions is strongly constrained by the energy conditions satisfied or violated by the model \eqref{Kondobraneaction}, hence the precise values of $\kappa_N,\mN,q,\mfC$ will not matter for the qualitative features of the model, and the above choice gives representative results. In any case, as argued above the model under consideration can serve as an interesting non-trivial AdS$_3$/BCFT$_2$ toy model.

\begin{figure}[htb]
	\centering
	\def\svgwidth{0.75\columnwidth}
	\executeiffilenewer{embeddingTurn_latexed.svg}{embeddingTurn_latexed.pdf}%
	{inkscape -z -D --file=embeddingTurn_latexed.svg %
		--export-pdf=embeddingTurn_latexed.pdf --export-latex}%
	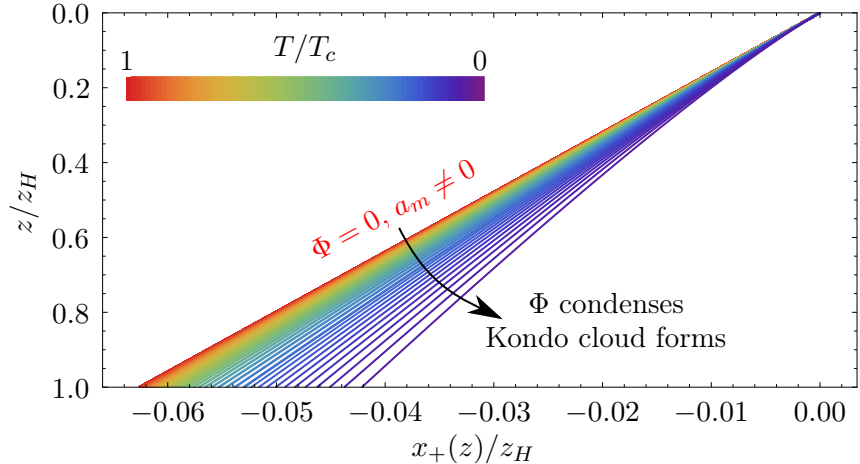%

	\caption[Brane embeddings for the holographic Kondo model]{Embedding profiles $x_+(z)$ for the embedding of the brane $Q$ into the bulk spacetime \eqref{BTZ}. Note that the bulk spacetime $N$ is located to the \textit{right} of the curves, at larger $x$-values. See also figure \ref{fig::NMQP} again. At $T=T_c$, the scalar field vanishes everywhere and the embedding is known to be given by 
		a constant tension solution \eqref{backgroundSolutionGaugeAndEmbedding}. As the temperature is lowered (or $\mu$ is increased), the scalar field condenses and the brane bends to the right. The figure is presented as in \cite{Erdmenger:2015spo}.
	}
	\label{fig::backreaction}
\end{figure}

The numerical results for the embeddings $x_+(z)$ of $Q$ into $N$ are shown in figure \ref{fig::backreaction}. As explained in section \ref{sec::Kondomodel}, in our numerics we make a coordinate choice that effectively keeps the event horizon radius $z_H\equiv1$ constant, so that it is strictly speaking $\mu$ that is varied. This has the benefit that the bulk spacetime $N$, given by the metric \eqref{BTZ}, stays the same. The effect of the backreaction is then that as we follow the RG flow, the brane $Q$ starts at its constant tension UV configuration and sweeps to the right over the fixed bulk spacetime like a curtain. It is hence immediately visible that the bulk spacetime will loose volume along the RG flow, in fact, we can in this setup directly identify specific bulk points that will be lost and determine the value of $T/T_c$ at which this will happen. 

Interestingly, we also see that for $z\rightarrow0$ all the curves in figure \ref{fig::backreaction} approach the boundary with the same slope. This is a simple consequence of the fact that the scalar field $\phi(z)$ falls off towards the boundary, hence as $z\rightarrow0$ the curves $x_+(z)$ will increasingly resemble the solution for the critical temperature where $\phi(z)=0$ everywhere. In RG flow parlance, if we interpret the near boundary region of the spacetime as UV region, it is clear that all curves along the RG flow should be similar in this region, as they were all derived by following the RG flow from the same UV fixed point. The consequence of this similarity of the embedding curves near the boundary is that when calculating the loss of volume for a certain point along the RG flow, we can expect the UV divergences near the boundary to cancel. We will now show this calculation explicitly. 

Following the proposal \eqref{16complexity}, we will define the \textit{relative complexity} via the proportionality\footnote{This definition is similar, but slightly different from the definition of a relative complexity given in \cite{Brown:2017jil} or the complexity of formation given in \cite{Chapman:2016hwi}.}        
\begin{align}
LG_N\times\mC_{rel}(T/T_c)\propto \mV^{T/T_c<1}-\mV^{T/T_c=1},
\label{Crel}
\end{align}
where $\mV^{T/T_c<1}$ is meant to be the volume of the (co-dimension one) $t=0$ slice of our bulk geometry for some $T/T_c<1$, while $\mV^{T/T_c=1}$ is the similar volume of the constant tension solution corresponding to the UV fixed point. Due to the time reflection symmetry of the BTZ geometry and its conformal diagram, the $t=0$ slice is an equal-time slice anchored at the two boundaries at times $t_L=t_R=0$ with extremal volume, as required by the prescriptions \eqref{16complexity} and \eqref{fidsus} \cite{Chapman:2016hwi}. It should be pointed out that while the definition of fidelity susceptibility can be easily generalised to the case of mixed boundary states \cite{MIyaji:2015mia,Bak:2015jxd,Trivella:2016brw,Banerjee:2017qti}, this is not so clear with complexity, see however \cite{Alishahiha:2015rta,Ben-Ami:2016qex,Carmi:2016wjl,Roy:2017kha}. We will hence assume that we are working in a two sided black hole spacetime, dual to a thermofield double (-like) pure state. As the numerical solutions obtained in \cite{Erdmenger:2015spo} and depicted in figure \ref{fig::backreaction} are only on one side of the Einstein-Rosen (ER) bridge (and outside of the event horizon), we have to conjecture that the state can be purified by adding another copy of the Kondo model on the other side of the ER bridge.\footnote{We do not do so without evidence: In fact, in \cite{Erdmenger:2015spo} it was shown that the embeddings depicted in figure \ref{fig::backreaction} approach a constant tension solution of the form \eqref{backgroundSolutionGaugeAndEmbedding} (however with a different value $s$) near the horizon, and it can be explicitly shown that such a constant tension solution can be analytically and symmetrically (!) extended throughout the entire Penrose diagram of the BTZ black hole, i.e.~behind the horizon and to the other side of the ER bridge. Furthermore, a similar purification of the finite temperature Kondo model was achieved in the Kondo MERA model proposed in \cite{Matsueda:2012gc}.}

In short, we define the relative complexity at $T/T_c<1$ to be proportional to the loss of bulk volume compared to $T/T_c=1$. The induced metric on the $t=0$ slice of the 
BTZ black hole \eqref{BTZ} reads
\begin{align}
ds^2=\frac{1}{z^2}\left(dx^2+\frac{dz^2}{1-\frac{z^2}{z_H^2}}\right),
\end{align}
and consequently we find
\begin{align}
\mV^{T/T_c<1}-\mV^{T/T_c=1}&= \int_{\epsilon}^{z_H}dz 
\frac{1}{z^2\sqrt{1-\frac{z^2}{z_H^2}}}\int_{x_+^{T/T_c<1}(z)}^{x_+^{T/T_c=1}(z)}dx
=\int_{\epsilon}^{z_H}dz 
\frac{x_+^{T/T_c=1}(z)-x_+^{T/T_c<1}(z)}{z^2\sqrt{1-\frac{z^2}{z_H^2}}},
\label{ComplexityIntegral}
\end{align}
where, for the moment, we have retained an explicit UV cutoff $\epsilon$. Although technically comparing the volumes of two different spacetimes, this formula only includes one parameter $z_H$, which as explained in section \ref{sec::Kondomodel} we effectively set to one for both spacetimes. We also use the same cutoff $\epsilon$ for the regularisation of the divergent volumes $\mV^{T/T_c<1}$ and $\mV^{T/T_c=1}$, allowing us to write the difference $\mV^{T/T_c<1}-\mV^{T/T_c=1}$ as an integral over a difference in \eqref{ComplexityIntegral}. We will discuss this detail further in section \ref{sec::conc}.

\begin{figure}[htb]
	\centering
	\def\svgwidth{0.65\columnwidth}
	\executeiffilenewer{relcomplex.svg}{relcomplex.pdf}%
	{inkscape -z -D --file=relcomplex.svg %
		--export-pdf=relcomplex.pdf --export-latex}%
	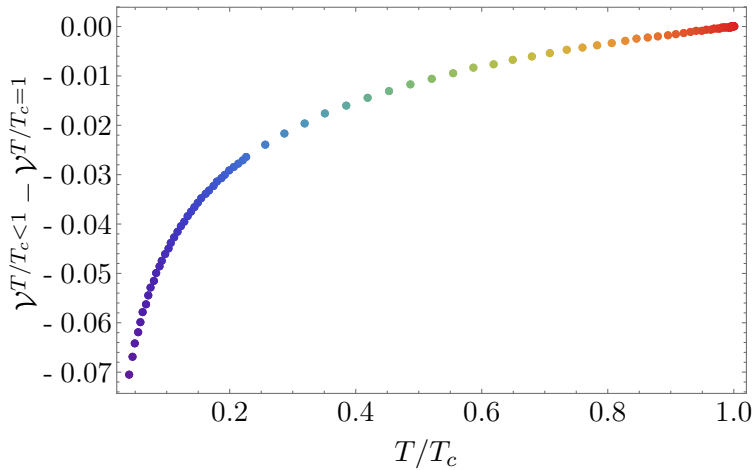%

	\caption[Proposal for relative complexity in the holographic Kondo model]{Relative complexity as defined in 
		\eqref{ComplexityIntegral} shown as a function of $T/T_c$. Although, as discussed in the text, in the Kondo model this 
		quantity is by definition finite as $\epsilon\rightarrow0$, we have calculated these points using an explicit cutoff 
		$\epsilon=10^{-10}$ in order to avoid numerical problems.  
	}
	\label{fig::relcomplex}
\end{figure}

The integrand in \eqref{ComplexityIntegral} may diverge both near the horizon $z=z_H$ and near the boundary $z=0$, however, the divergence at the horizon is mild and can be integrated over. What about the boundary? A priori, \eqref{ComplexityIntegral} might be divergent in the limit $\epsilon\rightarrow0$, but as we 
have discussed above the branes all approach $z=0$ with the same leading (linear) 
behaviour $x'^{(T/T_c<1)}_+(0)=x'^{(T/T_c=1)}_+(0)$, and as by definition also $x^{T/T_c<1}_+(0)=x^{T/T_c=1}_+(0)=0$, we see that $x^{T/T_c=1}(z)-x^{T/T_c<1}(z)$ goes to zero strictly faster than $const.\times z$. Hence, also the divergence near the boundary of the integrand of \eqref{ComplexityIntegral} will be mild enough to carry out the integral, yielding a finite result in the limit $\epsilon\rightarrow0$, see figure \ref{fig::relcomplex}.   

Let us briefly comment on the results depicted in figure \ref{fig::relcomplex}. First of all, as explained above, the depicted values are manifestly finite. Secondly, we see that as $T/T_c\rightarrow0$, the loss of bulk volume is monotonic. This may suggest that there is a complexity/fidelity susceptibility analogue of the Affleck-Ludwig $g$-theorem \eqref{lngT} for holographic BCFTs. We will discuss this possibility in more detail in sections \ref{sec::theorem} and \ref{sec::conc}. 

\subsection{Comparison to impurity entropy}
\label{sec::Simp}

Before doing so, however, it will be very instructive to compare the relative complexity defined in \eqref{ComplexityIntegral} to a quantity named \textit{impurity entropy}\cite{1742-5468-2007-01-L01001,1751-8121-42-50-504009,1742-5468-2007-08-P08003,
	PhysRevB.84.041107}:
\begin{gather}
S_{imp}(\ell)\equiv S(\ell)\big|_{\text{Impurity 
		present}}-S(\ell)\big|_{\text{Impurity absent}} \, ,
\label{Simpdef}
\end{gather}
where $S(\ell)\big|_{\text{Impurity present}}$ is the entanglement entropy for a boundary interval $[0,\ell]$ in the backreacted geometry depicted in figure \ref{fig::backreaction}, while $S(\ell)\big|_{\text{Impurity absent}}$ is the similar result for a trivial embedding $x_+(z)\equiv0$. Details on the calculation of this quantity in the holographic Kondo model were given in \cite{Erdmenger:2015spo}, here we will only quickly point out three relevant points:
\begin{itemize}
	\item Just like $\mC_{rel}(T/T_c)$, $S_{imp}(\ell)$ is manifestly finite as the UV divergences in \eqref{Simpdef} cancel. However, this cancellation is much less intricate, $S_{imp}(\ell)$ would remain finite even if $x'^{(T)}_+(0)=x'^{(T=T_c)}_+(0)$ would not hold in the Kondo model. Hence the finiteness of $\mC_{rel}(T/T_c)$ is a stricter consistency condition on the model than the finiteness of $S_{imp}(\ell)$. We will come back to this issue in section \ref{sec::conc}.
	
	\item As discussed in \cite{Erdmenger:2015spo}, this quantity can be related to the boundary entropy or \textit{$g$-function} via the limit 
	\begin{align}
	\ln(g)\equiv  S_{imp}(\ell\rightarrow\infty).
	\label{limit}
	\end{align}
	In fact, in \cite{Erdmenger:2015spo} it was explicitly checked that the holographic Kondo model satisfies the $g$-theorem \eqref{lngT} as expected for holographic BCFTs based on the proof by Takayanagi \cite{Takayanagi:2011zk}. This $g$-theorem is our main motivation to search for a similar monotonicity theorem for $\mC_{rel}$ in the next section. 
	
	\item It was shown in \cite{1742-5468-2007-08-P08003} that in the Kondo model at low temperatures
	\begin{align}
	S_{imp}=\frac{\pi^2c\,\xi_{\text{K}} T}{6v}\coth\left(\frac{2\pi\ell 
		T}{v}\right)
	\text{\ \ for\ \ }T\xi_\text{K}/v, \; \xi_{\text{K}}/\ell\ll1,
	\label{Affleck}
	\end{align}
	where $\xi_\text{K}$ is the \textit{Kondo length scale} and $v$ is the Fermi velocity. This result is not dependent on the specific details of the Kondo effect and can hence be expected to be valid for a broad class of BCFTs. Correspondingly, based on a simple geometrical approximation we showed in \cite{Erdmenger:2015spo} that this type of formula can also be reproduced very generically in holographic AdS$_3$/BCFT$_2$ models. To our current knowledge, no similar generic low temperature approximation formula exists for complexity or fidelity susceptibility of BCFTs, and no generic result for $\mC_{rel}$ analogous to \eqref{Affleck} can be derived in AdS/BCFT with simple geometrical approximations. This again shows that the quantity $\mC_{rel}$ is much more sensitive to the details of the model than $S_{imp}(\ell)$.    
	
\end{itemize}

\section{A holographic complexity/fidelity susceptibility
	\\
	analogue of the $g$-theorem}
\label{sec::theorem}

In this section we will prove the finding of section \ref{sec::volumeloss}, i.e.~the monotonic decrease of the bulk volume when moving along the RG flow, for general AdS$_3$/BCFT$_2$ models of the type \eqref{Action}. In oder to do so we need to clearly state the underlying assumptions of the proof that we are about to lay out. The assumptions will be that we are investigating a static  AdS$_3$/BCFT$_2$ model of the type \eqref{Action}, where the fixed bulk spacetime $g_{\mu\nu}$ is given either by a BTZ black hole or a \Poincare background \eqref{BTZ}\footnote{While this restriction technically leads to a loss of generality, due to the issues explained in footnote \ref{bulkflow} it seems reasonable to avoid any system that undergoes a flow of the central charge $c$ of the BCFT. For example, the bulk metric $g_{\mu\nu}$ should certainly not correspond to a domain wall solution interpolating between vacua with different effective AdS scales as in \cite{Freedman:1999gp}.  }. Furthermore, we assume that this model, similarly to the Kondo model of sections \ref{sec::Kondo} and \ref{sec::volumeloss}, displays an RG flow where instead of (or in addition to) the holographic coordinate $z$, we have some parameter $\mu$ that we can interpret to parametrise an RG flow, i.e.~where $\mu$ can be continuously varied between values $\mu_{UV}$ and $\mu_{IR}$ such that the embedding of $Q$ into the bulk spacetime is a function of $\mu$.

Additionally, we will need to make use of energy conditions on the energy-momentum tensor $S_{ij}$. The importance of such energy conditions in static AdS$_3$/BCFT$_2$ models has been discussed in great detail in \cite{Erdmenger:2014xya}. Let us briefly summarise: As the worldvolume of $Q$ is $1+1$ dimensional in an AdS$_3$/BCFT$_2$ model, it follows that in the static case, by effectively using lightcone-coordinates, we can decompose the energy momentum-tensor on $Q$ in terms of two scalar quantities $S$ and $S_{L/R}$ according to
\begin{align}
S_{ij}=\frac{S}{2}\gamma_{ij}+S_{L/R}\tilde{\gamma}_{ij}.
\end{align}
Here, $S$ is the trace of $S_{ij}$ while $S_{L/R}$ is the \textit{traceless part}. Similarly, $\gamma_{ij}$ is the induced metric as before, while $\tilde{\gamma}_{ij}$ is a symmetric traceless tensor that can be uniquely constructed from $\gamma_{ij}$ and the timelike Killing vector present by the assumption of staticity. In terms of these scalar quantities, various energy conditions take a remarkably simple form, e.g.
\begin{align}
\text{null energy condition (NEC):}&\ \ S_{L/R}\geq 0,
\label{NEC}
\\
\text{weak energy condition (WEC):}&\ \ 2S_{L/R}-S\geq 0,
\label{WEC}
\end{align}
and what we will refer to as the $1+1$ dimensional analogue of the
\begin{align}
\text{strong energy condition (SEC):}&\ \ 2S_{L/R}+S\geq 0.
\label{SEC}
\end{align}
While NEC and WEC are taking the same role in this study as in many other investigations of gravitational physics, it is important to reiterate the phenomenological importance of the SEC in AdS$_3$/BCFT$_2$ models that has been discussed in more detail in \cite{Erdmenger:2014xya,Erdmenger:2015spo}. In \cite{Erdmenger:2014xya}, we showed that when both WEC and SEC are non-trivially satisfied at the same time, the embedding profile $x_+(z)$ solving the equations \eqref{EOM} will return to the boundary in a $\cup$-shaped way. Conversely, in order to obtain embeddings for $Q$ which reach from the boundary to the event horizon as for example in figure \ref{fig::backreaction}, either WEC or SEC have to be violated. Indeed, in \cite{Erdmenger:2015spo} we found that the matter content \eqref{Kondobraneaction} in the Kondo model explicitly leads to a violation of the SEC. 

\begin{figure}[htb]
	\centering
	\def\svgwidth{0.59\columnwidth}
	\executeiffilenewer{delta_mu.svg}{delta_mu.pdf}%
	{inkscape -z -D --file=delta_mu.svg %
		--export-pdf=delta_mu.pdf --export-latex}%
	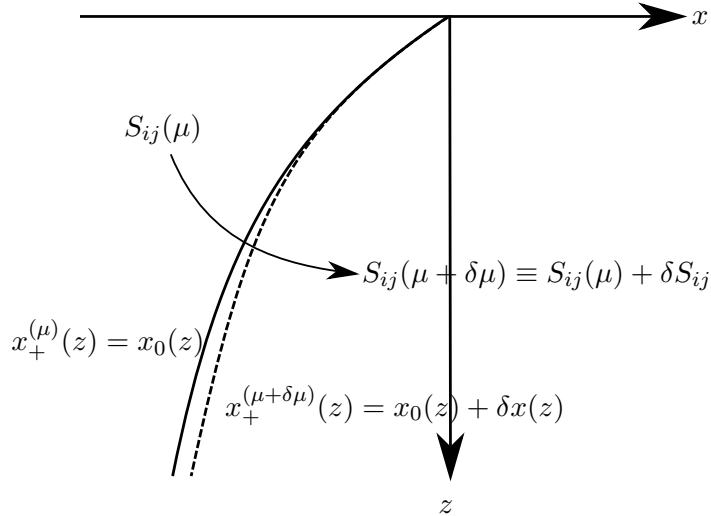%

	\caption[]{Geometric setup for the proof of volume loss along a boundary RG flow parametrised by $\mu$.
	}
	\label{fig::deltamu}
\end{figure}

In terms of energy conditions, we will hence make the following assumptions, see also figure \ref{fig::deltamu}. We assume that for some $\mu$ along the RG flow ($\mu_{UV}\leq\mu<\mu_{IR}$) the embedding is given by a function $x_+^{(\mu)}=x_0(z)$ solving \eqref{EOM} for the energy momentum tensor $S_{ij}(\mu)=S^0_{ij}$. Moving an infinitesimal step $\delta\mu$ along the RG flow, we find the embedding 
\begin{align}
x_+^{(\mu+\delta\mu)}=x_0(z)+\delta x(z)
\label{deltaX}
\end{align}
as the solution to \eqref{EOM} for the energy-momentum tensor $S_{ij}(\mu+\delta\mu)=S^0_{ij}+\delta S_{ij}$. For the physical reasons explained above, it seems prudent to assume that the NEC is satisfied and that the SEC is violated or saturated for both $\mu$ and $\mu+\delta\mu$:
\begin{align}
S_{L/R}\geq 0,\ S_{L/R}+\delta S_{L/R}\geq0,\ 2S_{L/R}+S\leq 0,\ 2S_{L/R}+2\delta S_{L/R}+S+\delta S\leq 0.
\label{ECassumptions1}
\end{align}
What conditions can or should we impose for physical reasons on $\delta S_{L/R}$ and $\delta S$ directly? In the Kondo model, along the RG flow the scalar field $\phi$ condenses and hence adds more and more positive energy to the energy-momentum tensor, consequently it seems sensible to demand the NEC
\begin{align}
\text{$\delta$NEC:}&\ \ \delta S_{L/R}\geq0
\label{deltaNEC}
\end{align}
for each infinitesimal step $\delta\mu$. Interestingly, in \cite{Erdmenger:2014xya} we found a close relationship between NEC and SEC, when expressed as functions of $z$. Energy-momentum conservation for a static embedding in a BTZ background \eqref{BTZ} implies
\begin{align}
\partial_z(S+2S_{L/R})=\frac{4}{z h(z)}S_{L/R}.
\label{conservation}
\end{align} 
As $h(z)\geq0$ outside of the horizon, this means that the NEC implies that the SEC has a tendency to become more satisfied (or less violated) as one moves from the boundary into the bulk. If the SEC is satisfied at the boundary and the NEC holds everywhere, then the SEC is also satisfied everywhere. This motivates that also along the RG flow parametrised by $\mu$, although the SEC will always be violated, it should have the tendency to become \textit{less violated} towards the IR\footnote{As explained before, fixed points are expected to be described by constant tension solutions. For such solutions, the NEC is saturated, and the tendency of the SEC to become less violated along the RG flow implies the statement that the tension of an IR fixed point will always be lower than the tension of a UV fixed point, but still nonnegative. This is also a requirement for the Affleck-Ludwig $g$-theorem to be satisfied.}. We will hence make use of the assumption\footnote{This assumption as well as \eqref{deltaNEC} are satisfied in the holographic Kondo model of section \ref{sec::Kondomodel}. Even more, in this model we find that the stronger condition $\delta S\geq0$ is also satisfied, which together with \eqref{deltaNEC} implies \eqref{deltaSEC}.}
\begin{align}
\text{$\delta$SEC:}&\ \ 2\delta S_{L/R}+\delta S\geq0.
\label{deltaSEC}
\end{align}

The next step is to insert \eqref{deltaX} into the equations of motion \eqref{EOM} and linearise for small $\delta x(z)$. Using \eqref{EOM} and \eqref{Ktensor}, we then find 
\begin{align}
\delta SEC\Rightarrow \frac{2 z_H^3 \delta x'(z)}{\left((z_H^2-z^2) x_0'(z)^2+z_H^2\right)^{3/2}}\geq0\Rightarrow \delta x'(z)\geq0
\label{SECapplied}
\end{align}
for all $0\leq z\leq z_H$. As the location of the defect $P$ is fixed by $x_+^{(\mu)}(0)=0$ for any $\mu$, \eqref{SECapplied} implies $\delta x(z)\geq0$ for any $0\leq z\leq z_H$. Analogously to \eqref{ComplexityIntegral}, we hence find\footnote{\label{FN::gtheorem} This is also related to the $g$-theorem. As explained in \cite{Erdmenger:2015spo}, due to \eqref{limit} $\ln g \propto -x_+(z_H)$, i.e.~the boundary entropy is given be the extra piece of black hole event horizon visible due to the non-trivial backreaction. The above argument then also implies the $g$-theorem \eqref{lngT}.} 
\begin{align}
\delta\mC_{rel}(\mu)\propto \mV^{\mu+\delta\mu}-\mV^{\mu}=\int_{\epsilon}^{z_H}dz 
\frac{-\delta x(z)}{z^2\sqrt{1-\frac{z^2}{z_H^2}}}\leq 0.
\label{deltaComplexityIntegral}
\end{align}
Hence we have proven that for any infinitesimal step $\delta\mu$ along the RG flow, the bulk volume cannot increase. Consequently, the bulk volume is a monotonically decreasing function along the RG flow parametrised by $\mu$, just as seen in the example of the Kondo model in figure \ref{fig::relcomplex}. It should be noted, however, that \eqref{deltaComplexityIntegral} will only be finite if $\delta x'(0)=0$, i.e.~if the $\delta$SEC \eqref{deltaSEC},\eqref{SECapplied} is \textit{saturated} at the boundary.\footnote{In the Kondo model of section \ref{sec::Kondomodel}, this is indeed the case for $T/T_c>0$ as discussed in \cite{Erdmenger:2015spo}. More generally, for a massive scalar field $\varphi$ living on the effectively $1+1$ dimensional space $Q$, we find $S\propto -M^2\varphi^2$ and $S_{L/R}\propto \partial^i\varphi\partial_i\varphi$. The $z\rightarrow0$ behaviour of the energy-momentum tensor on $Q$ will hence depend on the asymptotic behaviour of $\varphi(z)$, and the conditions \eqref{deltaNEC}, \eqref{deltaSEC} will be saturated at the boundary if no new modes of $\varphi$ that do not sufficiently vanish at the boundary get turned on as the parameter $\mu$ is varied. Note that in the Kondo model, the gauge field $a_t$ has a divergent component near the boundary in \eqref{Qmuc}, but as $\mQ$ is kept fixed, this does not affect the change of $S$ and $S_{L/R}$ near the boundary as $\mu$ is varied.}

\section{Discussion}
\label{sec::conc}

In this final section, we will discuss the results obtained in sections \ref{sec::volumeloss} and \ref{sec::theorem}, comment on a few technical details, and give an outlook on further possible interesting research directions.

As explained in section \ref{sec::Kondomodel}, when working with the Kondo model we have scaled $z_H\rightarrow 1$, so that the bulk metric $g_{\mu\nu}$ on $N$ in constant along the RG flow parametrised by $\mu$, and the change of the geometry manifests itself only in a change of the embedding $Q$ into $N$. However, we have argued that this should be equivalent to keeping $\mu$ fixed and varying $T/T_c$ instead. There now appears a technical detail relating to the choice of cutoffs. In the definition \eqref{Crel}, we are comparing two different spacetimes with each other, so strictly speaking we make \textit{two} choices of a cutoff, $\epsilon^{(T_c)}$ and $\epsilon^{(T)}$. In section \ref{sec::volumeloss}, we have chosen these two cutoffs to be the same \textit{after} scaling $z_H\rightarrow1$. If we had for example done so \textit{before} doing this rescaling, \eqref{ComplexityIntegral} would have gained an additional term
\begin{align}
\int_{\epsilon^{(T)}}^{\epsilon^{(T_c)}}dz 
\frac{-x_+^{T}(z)}{z^2\sqrt{1-\frac{z^2}{z_H^2}}}\approx -x_+'^T(0) \log\left(\frac{T_c}{T}\right)>0
\label{cutoffdependence}
\end{align}
which might spoil the decrease and monotonicity behaviour found in sections \ref{sec::volumeloss} and \ref{sec::theorem}. How serious is this problem, and how can we arrive at a physical choice of cutoffs $\epsilon$?\footnote{The appropriate choice of cutoffs when comparing the complexities of two different spacetimes was also discussed in \cite{Chapman:2016hwi} by choosing an asymptotic Fefferman-Graham expansion for both spacetimes. We will in the following present an argument that is more focused on the physical interpretation of the Kondo model or similar AdS/BCFT models. } To answer this question, we should first note that precisely the same problem appears in the definition of impurity entropy given in \eqref{Simpdef}. There, two divergent quantities calculated in two a priori different spacetimes are subtracted to obtain a finite result. Entanglement entropy for $1+1$ dimensional (B)CFTs diverges with $\sim \log(\epsilon)$ in the UV cutoff, so we see that a different choice of cutoff for the two terms in \eqref{Simpdef} would lead to a similar term as in \eqref{cutoffdependence}, spoiling for example the result \eqref{Affleck}. The question should hence be: \textit{Which} two states are we comparing in the definitions \eqref{Simpdef} and \eqref{Crel}, respectively? In the definition \eqref{Simpdef}, both states "Impurity present" and "Impurity absent" have physical meaning at any $T$. In principle it should be possible to prepare them in a lab and carry out measurements on them. The UV cutoff $\epsilon$ would then be physically set by the lattice spacing of the metal under investigation, which would indeed be the same for both cases. For the example of relative complexity \eqref{Crel}, the situation is a little bit more difficult: We need to specify what we precisely mean with the quantities labeled by $T=T_c$ in \eqref{Crel} and \eqref{ComplexityIntegral}. Clearly, the high temperature UV fixed point of the system cannot be prepared in a lab for low temperatures, as this would be oxymoronic. However, the reader should be reminded (section \ref{sec::Kondomodel}) that in the holographic Kondo model, the formation of the Kondo cloud is described by a phase transition (see footnote \ref{FN::largeN}). The UV fixed point at $T=T_c$ is described by the constant tension solution \eqref{backgroundSolutionGaugeAndEmbedding} with the tension given by \eqref{backgroundSolutionGeodesicLength}, while for $T<T_c$ the condensed phase is thermodynamically preferred. However, even for $T<T_c$ it is still possible to define the uncondensed phase and calculate the corresponding (constant tension) bulk geometry \eqref{backgroundSolutionGaugeAndEmbedding}. It may be unstable due to thermodynamic fluctuations, but at least in a mathematical sense it exists for any $T\leq T_c$, and can be used in \eqref{Crel} as the reference bulk spacetime in a physical way. As we are then in \eqref{Crel} genuinely comparing two spacetimes at the same temperature, it appears physical to choose the cutoffs to be equivalent.\footnote{A clever choice of reference states was also a part of the argument in \cite{Casini:2016fgb} leading to an entropic $g$-theorem.} \footnote{More generally, fixed points are described by constant tension solutions, with a specific value of the tension. However, in any BTZ background, irrespectively of the value of the temperature $T$, a constant tension solution with the same value can be constructed via a geodesic normal flow \cite{Erdmenger:2014xya}. This generically allows to define a unique analogue of the UV fixed point for any temperature. }   

As we have noted in section \ref{sec::volumeloss}, the finiteness of $\mC_{rel}$ is tied to much stricter conditions than the finiteness of $S_{imp}$. Furthermore, on the example of the low temperature approximation \eqref{Affleck} and the lack of a complexity analogue thereof, we saw that complexity is more sensitive to details of the model at hand. This can be seen both as a detriment and as an advantage. First of all, it might be a detriment, because holographic studies are usually most relevant to real world physics when focusing on universal quantities, behaviours or mechanisms. This is due to the simple fact that there is no real world physical system with an exact holographic dual. The absence of universal results for complexity that a broad range of BCFTs can be expected to share (like \eqref{Affleck} for impurity entropy) is hence disappointing. Also, the fact that the volume in \eqref{ComplexityIntegral} obtains non-negligible contributions both from the near horizon and the near boundary part of the spacetime seems to imply that $\mC_{rel}$ mixes up UV and IR contributions, and may hence not be a clean variable to study RG flows. However, the peculiarities of $\mC_{rel}$ may also have their advantages. As noted in section \ref{sec::volumeloss}, the reason why $\mC_{rel}$ is finite is that all curves approach the boundary with the same slope $x'_+(0)$. Using $z$ again as the RG scale instead of $T/T_c$ or $\mu$, this is clearly a consequence of all these curves coming from the same UV fixed point, so to say. The finiteness of $\mC_{rel}$ hence acts as a built in check whether the spacetimes compared in \eqref{Crel} really can belong to the same RG flow. Only once the system has \textit{fully} reached its IR fixed point, we expect the corresponding embedding to be given by a constant tension solution similar to \eqref{backgroundSolutionGaugeAndEmbedding}, however with a different tension, i.e.~a different (lower) value of $s$. Only then would $\mC_{rel}$ diverge. For the Kondo model, this would imply the expectation that as $T/T_c\rightarrow0$, the results in figure \ref{fig::relcomplex} would stay finite for finite $T$, but diverge at $T=0$. This would be an important qualitative difference between $\mC_{rel}$ and the boundary entropy $\ln g$, which is also monotonic but does not diverge even at $T=0$.  

The main result of this paper was the proof of volume loss in section \ref{sec::theorem}. If the holographic proposals \eqref{16complexity} and/or \eqref{fidsus} are correct, this would suggest the existence of a complexity/fidelity susceptibility analogue of the $g$-theorem. What would the physical interpretation of such a theorem be? In terms of \eqref{16complexity}, a monotonic decrease of complexity along the RG flow would indicate that as we flow from the UV to the IR, the state of the field theory gets simpler. A hypothetical quantum-computer would need less gates to prepare the IR state from a simple reference state than to prepare the UV state. Qualitatively, this seems like a reasonable physical statement, and the theorem of section \ref{sec::theorem} makes this more precise in a certain way. Interestingly, as we have speculated above, in the holographic Kondo model it is expected that $\mC_{rel}$ stays finite along the RG flow but likely diverges for $T\rightarrow 0$. This would mean that taking a generic state from along the RG flow, it would be possible to undo the RG flow and recreate the UV state with a finite number of quantum gates. Taking the IR state however, from which \textit{all} UV degrees of freedom have vanished, it would take us an infinite amount of quantum gates to recreate the UV state. It will be interesting to compare this intuition with results from tensor network models of BCFTs \cite{Matsueda:2012gc,Czech:2016nxc}, or the entropic $g$-theorem proven in \cite{Casini:2016fgb}. There, the RG flow was interpreted in quantum information theoretic terms as an increasing \textit{distinguishability} between the UV state and states along the RG flow. Qualitatively, the more quantum gates a quantum computer needs in order to transform one state into the other, the more distinguishable the two states might be. The $g$-theorems of this paper and of \cite{Casini:2016fgb} hence seem to point in a similar direction, although they are very different in their details. Furthermore, if the conjecture of \cite{Brown:2017jil} that the lack of complexity (called \textit{uncomplexity}) is a resource in quantum computations is correct, then general results about when systems tend to attain low complexity, such as a complexity $g$-theorem, may be very useful. 

From the point of view of the proposal \eqref{fidsus} on the other hand, our results suggest the following qualitative intuition: The fidelity susceptibility studied in \cite{MIyaji:2015mia} measures the closeness of two states after a perturbation by a marginal operator. The monotonicity theorem of section \ref{sec::theorem} hence implies that as we move towards the IR, the field theory states (even after a small perturbation) become more and more similar. In the IR, so to say, the perturbed states are still closer to each other.

Last but not least, there are a few future research directions that we can now propose. Can the proof of section \ref{sec::theorem} be formulated with a less restrictive set of assumptions? Can one compare the geometrical results of this paper and of \cite{Erdmenger:2015spo} to tensor-network (specifically MERA) models of BCFTs \cite{Matsueda:2012gc,Czech:2016nxc}? After all, complexity by its very definition is clearly a quantity of relevance for tensor networks. Can we generalise the results of this paper to higher dimensional AdS/BCFT models? This may not be trivial, as in \cite{Chapman:2016hwi}, it was found that for AdS$_{2+1}$ the complexity of formation had some rather peculiar features that did not generalise to higher dimensions. Also, many AdS$_3$/BCFT$_2$ results utilising energy conditions do not generalise easily to higher dimensions as noted in \cite{Erdmenger:2014xya}.  Is there also a complexity/fidelity susceptibility analogue of the holographic $c$-theorem proven in \cite{Freedman:1999gp}? A number of complications in this work came from the fact that following the $g$-theorem \eqref{lngT}, we used $T/T_c$ respectively the chemical potential $\mu$ as parameter along the RG flow, instead of the radial coordinate $z$. This allowed us to avoid working with subregion complexity, for which a field theory definition in terms of a number of quantum gates is much less clear, due to the dual state being mixed. See however the discussions in \cite{Alishahiha:2015rta,Ben-Ami:2016qex,Carmi:2016wjl,Roy:2017kha}. Would the use of subregion complexity allow us to derive a different kind of $g$-theorem, maybe more similar to the zero-temperature "entropic $g$-theorem" of \cite{Casini:2016fgb}? The most important question is: Can we derive a similar monotonicity theorem for the action proposal \cite{Brown:2015bva,Brown:2015lvg} of complexity? In the AdS/BCFT models studied in this paper, the loss of volume $\mV$ was not a result of some intricate geometric feature affecting only the extremal slice at $t=0$, it was simply the result of a loss of bulk points. This implies that generically, lengths and volumes, however they may be precisely defined, will tend to decrease along the RG flow. In this sense, we can also view the $g$-theorem \eqref{lngT} as a consequence of the volume loss, as it is satisfied because certain bulk geodesic lengths get shorter along the RG flow \cite{Erdmenger:2015spo}, see also footnote \ref{FN::gtheorem}. Hence, we have reason to expect that the Wheeler-DeWitt (WDW) patch will also shrink. In the action proposal \cite{Brown:2015bva,Brown:2015lvg}, this WDW patch is the co-dimension zero volume over which the action is to be integrated. Consequently, if the integrand does not grow strong enough to compensate the shrinking of the integral domain, it is indeed expected that a $g$-theorem analogue will also hold for the action proposal for holographic complexity. The detailed study of these questions will be left for future research, however.

\section*{Acknowledgements}

The author is grateful to Dongsu Bak, Andrew O'Bannon, Adam Brown, Dean Carmi, Shira Chapman, 
Romuald Janik, Hugo Marrochio, Max-Niklas Newrzella, \'Alvaro V\'eliz-Osorio, Sotaro Sugishita and Tadashi Takayanagi for discussions and comments. The author was supported by the NCN grant 2012/06/A/ST2/00396.







\providecommand{\href}[2]{#2}\begingroup\raggedright\endgroup

\end{document}